# Graphene Aerogel Ink for the Inkjet Printing of the Micro-Supercapacitors

[1]Anand P S Gaur, [1]Wenjun Xiang, [1]Ping-Ping Chen [2]Arjun Nepal, [3]Brice Lacroix, [2]C M Sorensen, [1]S R Das

[1]Department of Industrial and Manufacturing System Engineering, Kansas State University, Manhattan, KS, 66506, USA

[2]Department of Physics, Kansas State University, Manhattan, KS, 66506, USA

[3]Department of Geology, Kansas State University, Manhattan, KS, 66506, USA

**The advances in the mass scale manufacturing of microscale energy storage devices via inkjet printing rely on the development of high-quality printable ink. The earth-abundant, non-toxic carbon materials such as graphene, carbon nanotube (CNT), reduced graphene oxide (r-GO) have shown excellent electrochemical performance and thus garnered significant interest as suitable electrode material. Here we report the formulation of printable graphene aerogel ink and the fabrication of the micro-supercapacitors (μ-SCs) on flexible polyimide substrates via inkjet printing method. The advantage of using pristine graphene aerogel intends to avoid the complex processing steps and the use of toxic chemicals in the ink formulation and lower the concentration of other additive components. Thus, a higher loading of active functional material in the printable ink is achieved. The aerogel ink directly employed to write the interdigitated μ-SCs devices on a flexible polyimide substrate at room temperature via inkjet printing. The electrochemical performance measured using the organic ionic liquid in the voltage range of 0-1 volt. These printed μ-SCs showed an areal capacity of 55 μF/cm$^2$ at a current density of 6 micro-amp/cm$^2$. The printed devices showed good stability, with ~80% of capacity retention after 10,000 cycles. Contrary to the graphene-based μ-SCs, the aerogel micro-supercapacitors do not show a significant distortion in the CV scan even at a very high scan rate of ~2Vs$^{-1}$. Thus, we propose graphene aerogel as promising electrode material for mass-scale production of the μ-SCs.**

**Introduction.**

The pervading flexible electronic devices are spanning its application from the user-inspired electronics, namely foldable display, energy storage[1],harvesting[2] etc. to the smart sensors for human health diagnostics and communication systems enabling the internet of things (IoT)[3–5]. Thus, there is an illustrious research thrust for novel functional materials and printing technology empowering the scalable fabrication of foldable devices. Among these, the small size, foldable energy storage devices are the critical aspect of this cutting-edge research making an electronic device portable in a true sense.[6,7] The ubiquitous lithium-ion batteries are mostly used in consumer portable electronics; however, the downsides to cite few, slow charging rate, poor cyclability, environmental safety concerns, and bulk size are holding back LIBs application in flexible devices.[8,9] Therefore, the active research work is persuaded in the manufacturing of the size compatible, rechargeable, flexible energy storage devices.[10–12]

Here, the electrochemical double-layer supercapacitors (EDLCs) are the most anticipated alternative of the conventional LIBs for the portable energy storage device. The charge in the EDLCs stored in the proximity of the polarized electrode surface by forming an electric double layer leading to the very high specific capacitance. The charging principle, as stated, evades the sluggish chemical reaction kinetics limiting the charging rate and life span of electrode material in LIBs. Thus, EDLCs can charge-discharge rapidly for thousands of active operating cycles without considerable corrosion of electrodes material and loss of specific capacity. The vertical geometry of EDLC comprising separator sandwiched between two active electrodes is prevalent for bulk production[6,13–16]; however, recent advances in solution phase printing technology exceedingly supported the in-plane IDEs architecture. The planar architecture is well suited for the direct printing of the micro size supercapacitor along with other functional electronic parts.

The planar architecture facilitates the lateral diffusion of ions during the charge-discharge process. Consequently, electrically conductive layered materials are the preferred for supercapacitor application.[13,17,18] Amongst these materials, graphene remains superior electrode material due to the high surface area ~ 2630 m$^2$/gm (theoretical), excellent electrical conductivity and superior chemical resilience.[19] Thus, to achieve the near theoretically estimated surface area to maximize the capacitance in graphene based supercapacitors, structurally engineered porous carbon[15,20], onion like carbon[14], porous graphene aerogel[21] used earlier for device fabrication. However, these processes are incompatible for scalable manufacturing, subsequently the facile methods such as liquid phase[22–24] or shear exfoliation[25] widely accepted as scalable production of high quality pristine graphene powder. The stable graphene suspension produced extensively from the natural graphite exfoliation in organic solvents where the use of polymer binders increased the yield comprehensively.

In an alternative approach, pristine graphene nano-sheets were produced at gram scale by the detonation of the hydrocarbons ($C_2H_2$) in a partial oxygen ($O_2$) environment.[26] The proposed method avoids the use of toxic chemicals, reported previously[24], and lengthy processing time. Thus, it could favor the industrial-scale manufacturing of the nano-graphene powder. The unique aspect of the graphene obtained in detonation method is the peculiar microstructure that contains graphene nano-sheets with the inherent nano size pores. The graphene nano-sheets eventually bonded together forming a gel like morphology. The nano porous channels could give access to greater surface area and hence higher energy storage could be achieved. Previously, a dramatic increase in electrochemical energy storage has been reported in 3D graphene aerogel by tuning the pore size.[15] However, this unique gel like formation of graphene nano sheets has not been utilized so far to formulate the printable ink energy storage devices printing. Thus, here we report the ink formulation protocol of the graphene aerosol-gel (Gr-AG) synthesized by the detonation method. The designed ink successfully implemented for the inkjet printing of the interdigitated micro-supercapacitors (MSCs) on the flexible polyimide sheet (25 microns thick). The printed supercapacitor showed a good areal capacitance and higher stability. Thus, the graphene aerogel ink formulation could offer an industrial-scale printing platform for the miniaturized energy storage device.

**Methods.**

**Materials characterization.** The microstructure of the GrAG studied by employing the Philips CM-100 transmission electron microscope at the accelerating voltage of 100 kV. The TEM specimen was prepared directly on the TEM copper grid by dipping the copper grid directly into the synthesized GrAG. Surface morphology and the uniformity of the printed electrode measured with Hitachi field emission scanning electron microscopy (FESEM). Raman (Renishaw Invia Raman Microscope, excitation wavelength 532 nm) and XPS spectrum ((PHI 5000 Versa Probe II, Physical Electronics Inc.) directly measured on the printed device to determine the phase and elemental analysis. The XPS spectrum was achieved with a combination of electron and argon ion flood guns. The X-ray beam size was 100 μm and survey spectra were recorded with pass energy (PE) of 117 eV step size 1 eV and dwell time 20 ms, whereas high-energy resolution spectra recorded with PE of 23.5eV, step size 0.05 eV and dwell time 20 ms.

**Aerogel ink preparation.** Graphene aerosol-gel (Gr-AG) powder obtained by the detonation method consists of pristine graphene nano-sheet agglomerates, thus to disperse the agglomerates probe sonication employed for 30 min in the ice bath condition with an ultrasonic probe (500 W, 20 kHz, Q500 sonicator, USA). 250 mg of Gr-AG powder dispersed in 50 ml ethanol and 1 w/v % ethyl cellulose (EC, Sigma-Aldrich, 4 cP grade measured in 80:20 toluene: ethanol at 5 wt%,

48% ethoxy) used as an emulsifier. The suspension then filtered through a 5-micron glass fiber syringe filter to remove the bigger chunks. The collected suspension then flocculated by adding the NaCl aqueous solution (0.04 gm/ml in deionized water) followed by vacuum filtration using a 0.45-micron nylon filter. The obtained GA/EC paste then dried at the hot plate at 70 ºC. The ink prepared by homogenously suspending the GA/EC powder in cyclohexanone and terpinol (volume ratio of 85:15) in bath sonication at a concentration of 70 mg/ml.

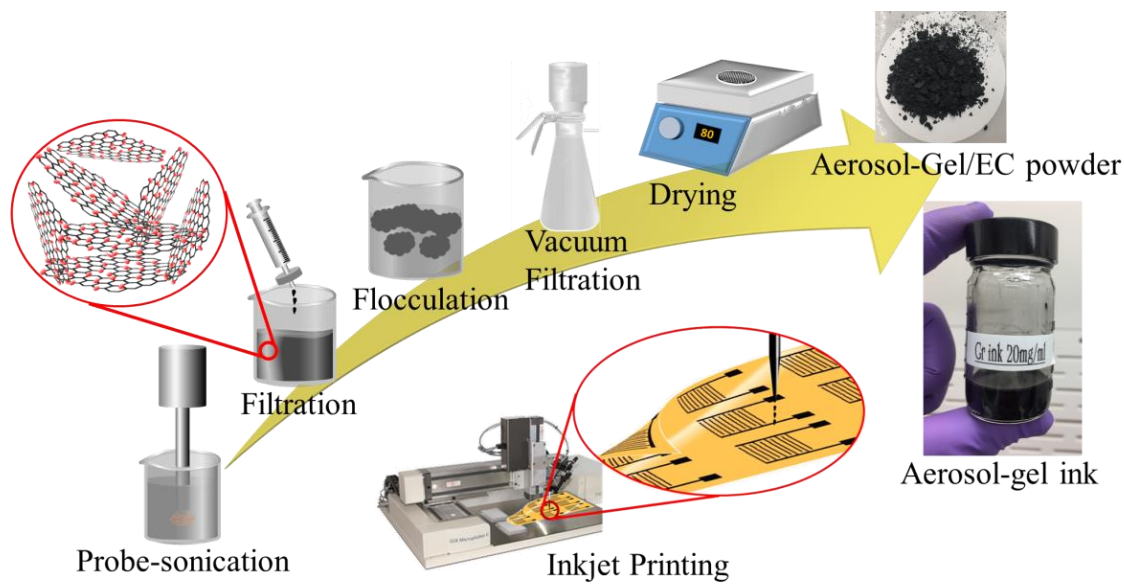

**Figure 1. Schematic illustration of the graphene aerosol-gel ink formulation and inkjet printing of the micro-supercapacitor device.** The upward arrow shows the ink formulation steps beginning from the dispersion graphene aerosol-gel powder to the dried ethyl cellulose/graphene aerosol gel powder. The formulated ink utilized to print micro-supercapacitors on polyimide substrate (25 microns).

**Inkjet printing of interdigitated electrodes (IDEs).** The interdigitated electrodes (IDEs) of MSCs were patterned on a flexible substrate by using the Ink-jet printer (SonoPlot, Microplotter II, USA) with a 20-micron nozzle size glass tip at room temperature. The substrates were thoroughly cleaned via bath sonication in the acetone and methanol mixture and dried by blowing off the nitrogen gas before printing. The printed electrodes were then heat-treated at 350 C for 2 hours in $N_2/H_2$ mixture (5% hydrogen in nitrogen) to burn off the organic binder.

**Electrochemical Performance.** Electrochemical performances of printed micro-supercapacitors were carried out by Gamry interface 1010 E potentiostat/Galvanostat in the potential window of 0-1 volts using the 1-Ethyl-3-methylimidazolium tetrafluoroborate (EMIM-BF4, Sigma Aldrich) organic electrolyte. The areal ($C_A$), volumetric ($C_V$) capacitance and equivalent series resistance ($R_{ESR}$) measured from the galvanostatic charge-discharge curve using the equation (1), (2) and (#)

$$C_A = \frac{I}{A\frac{dv}{dt}} \quad . \tag{1}$$

$$C_V = \frac{I}{V\frac{dv}{dt}} \tag{2}$$

$$R_{ESR} = V_{drop}/2I \tag{3}$$

Where the parameters I, A and V are the applied current, total area of the printed electrode fingers, total volume of the fingers respectively. The dv/dt is the slope of the discharge curve and $V_{drop}$ is the voltage drop at the beginning of the discharge cycle.

**Results and discussions.**

In past, colloidal chemistry employed to produce stable CNT and graphene suspension[27], and later this approach adapted successfully to produce the stable graphene flakes suspension in the organic solvents utilizing ultrasonic probe.[28] The use of surfactant such as ethyl cellulose (EC), nitro cellulose (NC) have shown significant impact in the yield of the graphene flakes. The surfactant encapsulated graphene flakes directly used to produce highly stable inks for the inkjet printing. We adapted the similar approach but with few modifications in the process as described in schematic illustration (figure 1) and experimental section of the ink preparation for the aerosol-gel powder to formulate the ink. The thermogravimetric analysis (TGA) carried out to measure the graphene content in the processed Gr-Ag/EC powder and showed in figure 2 (a). The change in mass as function of temperature denoted the decomposition of the surfactant at onset temperature ~250 °C. The significant change in the mass (30 wt %) occurred from 250 °C to 350 °C that indicate the complete decomposition of the surfactant into aromatic compounds.[29,30] Thus the Gr-AG/EC powder consists 70 wt % of graphene. It is worth to bring in notice that the weight percent of the graphene in present work is higher than reported earlier.[31,32] The Gr-AG/EC powder suspended further in cyclohexanone/terpinol mixture (85:15 volume ratio) to prepare the ink for the inkjet printing.

The formulated ink showed prolonged stability and printability. Figure 2 (b, c and d) represents the optical image of the inkjet-printed wild cat logo, interdigitated µ-SC and resistive elements respectively, on the flexible polyimide (25-µm thick) substrate. Geometrical dimensions of the printed device have been provided in the supplementary information (SI-I). To determine the homogeneity of the printed pattern we recorded the SEM and AFM image (SI-II), and it can be seen clearly from figure 2 (e) that printed patterns are highly uniform and free from the coffee ring effect. Thus, during the printing process ink flow was consistent and free of agglomeration resulting the good rheological properties of the formulate ink.[31,32] We scrutinized the printability of the ink by printing multiple devices with a double number of fingers (See image in SI-III). All the printed patterns have shown good homogeneity. Further, we recorded a high-resolution image of the printed electrode surface by SEM to study the surface morphology. The surface of the printed electrode comprised of spherical particles, forming a highly porous surface. Such porous morphology favors the storage capacity; however, it could have an adverse effect on the electrical conductivity of the printed patterns.

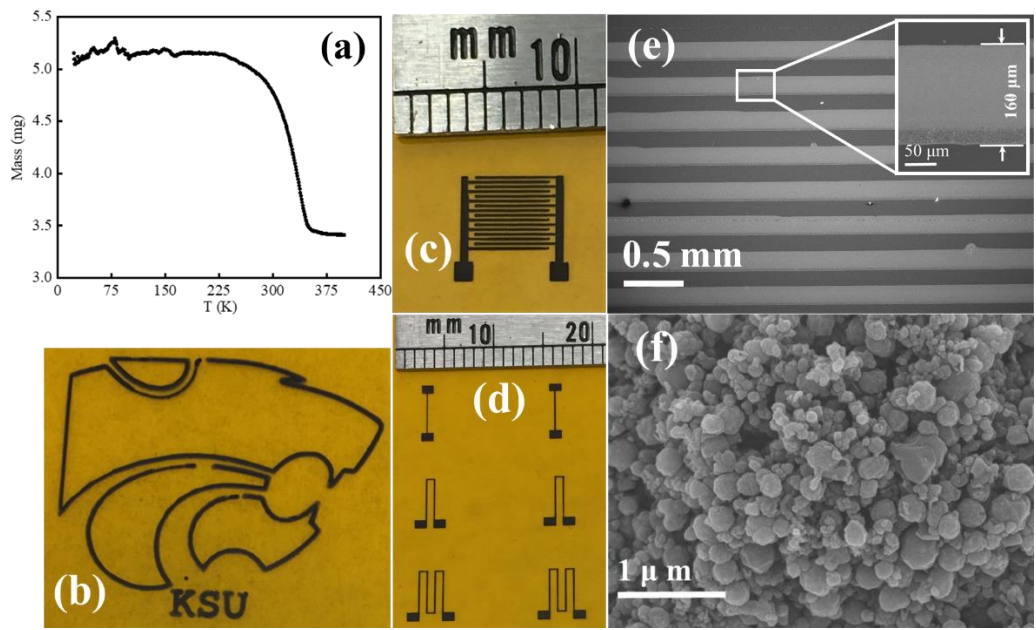

**Figure 2.** (a) TGA of Gr-AG/EC composite powder showing change in mass with temperature. (b, c and d) Optical image of the printed wild cat logo, interdigited μ-SC, and resistive elements. (e) SEM image of the inkjet-printed fingers. Inset figure shows the higher magnification of one such figure demonstrating the width and uniformity. (f) The surface morphology of the printed finger showing sphere like particles.

**Microstructural characterizations**. The microstructure of the Gr-AG powder represented in high-resolution TEM measurements, shown in figure (3). Figure 3 (a) shows the aggregates of graphene nano-sheets (GNS) with a near-uniform size distribution close to 100 nm. It is noteworthy that the edges of the GNS have darker contrast, as shown in figure 3 (b), with respect to the center of the sheet. As evident from the high magnification image of figure 3 (b), that the boundary of the GNS basically contains a shell-like structure where the shell is formed by edge terminated graphene sheets, as shown in the inset image of figure 3 (c). The randomly oriented sheets merge to form inherent nano porous system i.e. aerogel like structure. Thus the - synthesized nano-size graphene powder via the detonation method could be termed as graphene aerosol-gel (Gr-AG). The abundance of the edge terminated graphene, as shown in figure 3 (d), resembles the morphology of the carbon onion, where the graphene sheets arranged in concentric fashion to form closed multi-shell structure.[14,33] In the present case, shell structure confined to the edges of the particle only thus the peculiar microstructure could benefit the electrochemical energy storage due to inherit porosity.

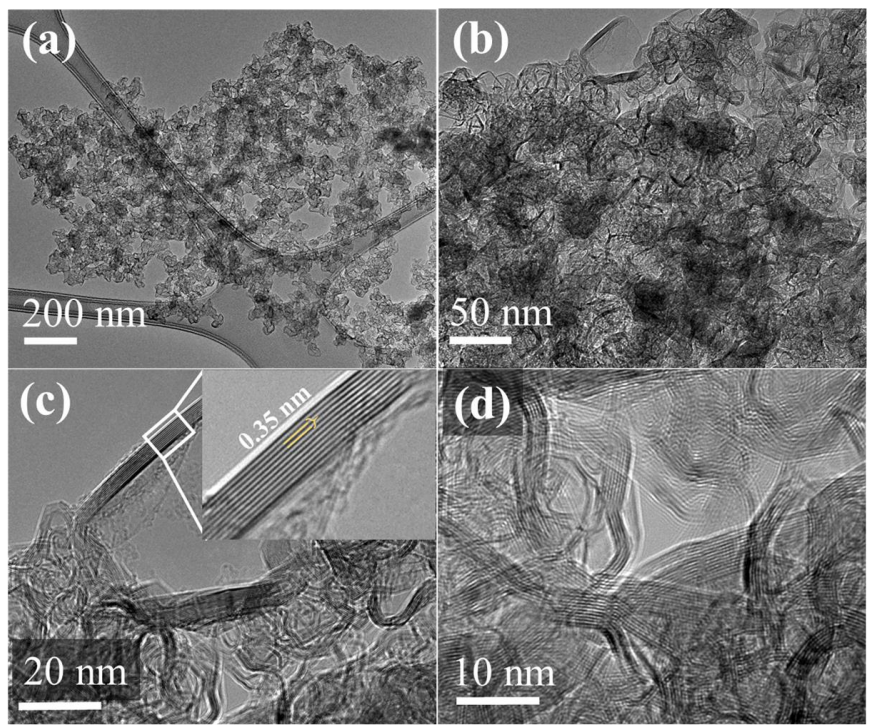

**Figure 3.** High-resolution TEM measurement for the microstructural characterization of graphene aerosol-gel. (a) The aggregate of graphene nano-sheets suspended on the TEM copper grid. (b) Higher magnification image of graphene nano-sheets demonstrating the higher contrast at the boundary with respect to the center of the sheets. (c) High-resolution image of one such nano-sheet showing the fringe/stripe-like microstructure at the boundary. The inset image shows that the stripes are basically formed by the terminating edge of the graphene sheet (d) Domination of stripe-like microstructure.

**Raman spectrum of graphene aerosol-gel**: Raman spectroscopy used in a persuasive manner as a nondestructive, high throughput characterization tool for the different sp² carbon materials. The unique band structure of graphene led to evolving the intense Raman bands due to resonant phonon scattering. Thus, the careful analysis of the Raman spectrum used to unveil significant microstructural aspects pertaining to defects, stacking order, number of layers, doping, stress and thermal conductivity of graphene. Figure 4 (a) represents the Raman spectrum of the graphene aerogel-sol recorded at room temperature. The Raman spectrum fitted with the Lorentzian function and contains three intense Raman bands centered at 1351 cm$^{-1}$, 1583 cm$^{-1}$ and 2700 cm$^{-1}$. These optical Raman active phonon modes are typically assigned as D, G and 2D band and attributed to $A_{1g}$, $E_{2g}$ and overtone of $A_{1g}$ phonon modes respectively. Additionally, two weak Raman bands at 1622 cm$^{-1}$ and 2452 cm$^{-1}$ are also present.

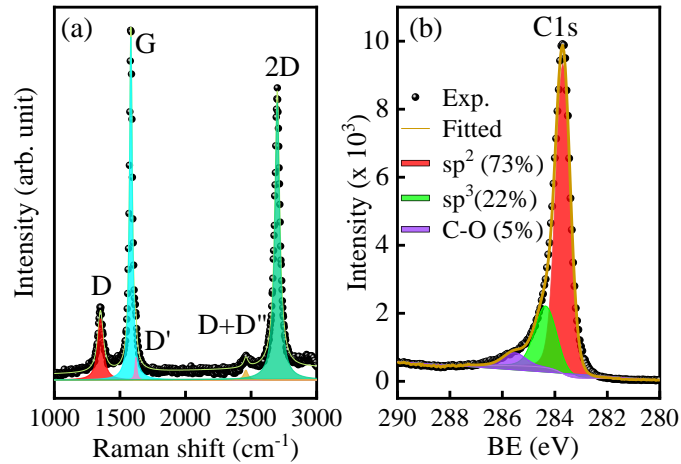

**Figure 4.** (a) Raman, and (b) C1s core level XPS spectrum of graphene GrAG.

These phonon modes are assigned previously as D′, D+D″ band respectively. The origin of the D′ is due to the intravalley double resonance (DR) scattering process around the K (or K′) point of the Brillouin zone. However, D+D″ mode activated from the combination of the LA branch phonon at 1100 cm$^{-1}$ and D phonon at K point of the Brillouin zone[34]. Form figure 4 (a), the emergence of the intense 2D band is the signature of graphitized carbon. Additionally, the line shape of the 2D band fitted very well with a single Lorentzian function, similar to the 2D band in the single-layer graphene (SLG). However, the upshift in wave number ~20 cm$^{-1}$ and higher full width at half maxima (FWHM) with respect to the SLG 2D band is consistent with turbostatic stacking of the graphene layers in the Gr-AG.

Moreover, the intensity of the D band is generally correlated with defect density exist in the form of structural defects, and disordered edges due to the loss of translation symmetry. The microstructure of aerosol-gel, as seen from the high-resolution TEM images, consists shell-like structure with the average size distribution of 100 nm. The boundary of the shell is confined by edge-oriented graphene sheets which persist substantially in the Gr-AG morphology. Thus, intuitively, a high magnitude of the I(D)/I(G) is expected in Gr-AG due to significant amount of exposed edges as shown previously in onion like carbon.[14] On the contrary, the lower magnitude of I(D)/I(G) ~0.2, implying that the not only the lower concentration of the structural defects, also the edges persevered the translation symmetry to a good extent. Thus, the edges have either zig-zag or arm chair ordering of the carbon atom with lower amount of edge defects. It has been shown by Cançado et al. that zig-zig edges do not contribute effectively to the D band intensity due to the conservation of momentum,[35] implying that the stripe like structure consists of substantial density of edge oriented graphene sheets with arm chair ordering of the carbon atoms.

**X-Ray photoelectron spectroscopy.** The chemical purity of the Gr-AG analyzed in XPS spectrum recorded from the surface of the printed device. The survey spectrum (SI-IV) showed only photo peaks relevant to C1s carbon and very low concentration of the oxygen. Figure 4 (b) represents the high-resolution scan of the C1s core level XPS spectrum of the Gr-AG. The asymmetric shape of the XPS band deconvolution into the three components. The greater intensity of the XPS band is shared by the sp² hybridized states (284.05 eV) stemming from the C=C of the hexagonal network of the carbon atoms[36] further confirmed the chemical purity of Gr-AG. The sp³ hybridize state (284.7 eV)[37] also exit in the XPS spectrum with fair concentration (22 %) along with minimal concentration of C-O groups.[11] It could be referred to the size and considerable amount of graphene edge states which are chemically more active to react with oxygen.

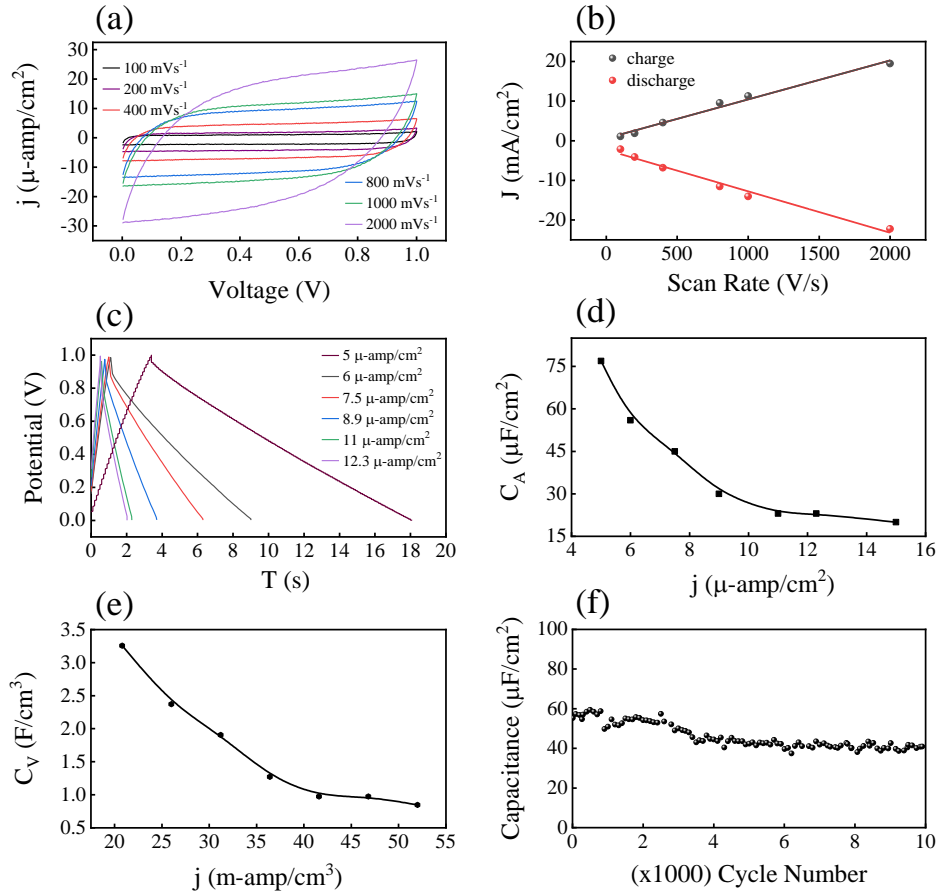

**Figure 5. Electrochemical performance of the printed μ-Sc.** (a) CV curves measured at different scan rates. (b) The linear increment of charge and discharge current (current density taken at the median of voltage window for charge and discharge currents) from the CV scans. (c) Galvanostatic charge-discharge curve measured at different current density. (d) The areal capacitance ($C_A$), and (e) volumetric capacitance measured for different current densities. (f) Cycling performance of the μ-SC device measured at 5 μ-amp/cm².

**Electrochemical performance**. The electrochemical performance of the printed Gr-AG μ-SC examined by the cyclic voltammogram (CV) at different scan rates, showed in figure 5 (a) in a potential window from 0.0 to 1.0 volt. The typical rectangular shape of the CV curves indicates the ideal double-layer capacitive characteristics of the printed µ-SC. The rectangular shape of the CV curve persisted in linear fashion as shown in figure 5 (b) for high scan rates measured up to ~2 V/s. However, the rounded corner (left top and bottom right) implied the significant magnitude of the equivalent series resistance (ESR) exist in the printed device. The ESR value calculated from the voltage drop ($\Delta V=36$ mV at 5 µ-amp/cm$^2$) at the beginning of the discharge curve using the equation (2). The magnitude of $R_{ESR}$ found to be ~45 k$\Omega$. The high magnitude of ESR raised from contact resistance, electrode-electrolyte interface resistance, bulk electrode resistance. It is apparent from the SEM image of the figure 2 (d) that the high porosity of Gr-AG electrode could have major contribution to the ESR magnitude. Further, the charge-discharge (CDC) profile measured at different current density of the MSC and showed in figure 5 (c). The CDC represents the typical triangular shape profile but with renowned asymmetrical shape particularly at lower current densities, however at higher current density triangular shape become more symmetrical. The areal ($C_A$) and volumetric ($C_v$) capacitance calculated from the slope of the galvanostatic discharge profile using equation (2), (3) and shown in figure 5 (d), (e) as a function of current density. The capacitive retention is poor of these devices, thus seek further investigation for better understanding of the Gr-AG devices to improve the device performance. The stability of the printed supercapacitor tested in an extended number of CDC cycles at a constant current density of 6 micro-amp/cm$^2$ and shown in figure 5 (f). The device showed good capacitance retention ~80% after 10,000 cycles. In order to increase the power density for the practical applications, generally multiple cells are assembled in series and parallel combination as shown in figure 6 (a),(d). The series combination, as anticipated, showed a decrease in capacity (figure 6 (b,c)) by a factor of three when operated in the voltage window of 0-1 volt, while showed a small increase in charge-discharge time operated up to 3 volts. Similarly, the parallel arrangement showed an increase in capacitance (figure (e) and (f)) by a factor of three compare to single cell. Thus formulated Gr-AG ink can be used directly to print the multiple devices in series and parallel combination directly in order to tailor the output.

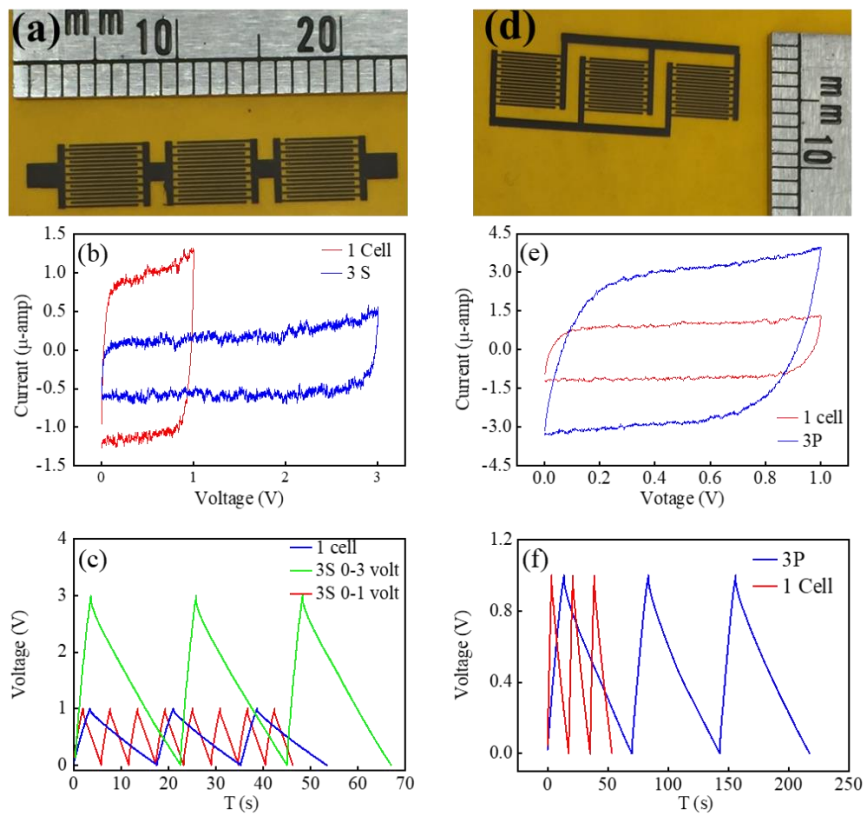

**Figure 6. Comparative study of the electrochemical response of the multiple device assembly.** (a) and (d) three printed μ-SC connected in series and parallel configuration respectively. (b) and (c) shows the CV scans recorded at 500 mV/s. (c), (f) shows the charge discharge profile of the assembled devices.

**Conclusions.**

In conclusion, it is the first ever report on electrochemical performance of the inkjet printed μ-SC derived from the Gr-AG. We developed the inkjet printable ink derived from the Gr-AG obtained by catalyst free detonation method of the hydrocarbon (acetylene) in partial pressure of oxygen. The derived ink use to print multiple μ-SC devices on flexible substrate. All the printed patterns showed high uniformity without any apparent stains or coffee ring effect. Thus, good printability and prolonged stability of the aerosol-gel ink confirmed the successful ink formulation protocol. The printed µ-SC showed good excellent capacity retention ~80% over extended number of charge-discharge cycles (10, 000 cycles) operated at 6 µA-cm$^{-2}$ in potential window 0-1 volt. Thus this approach could pave the gap of mass production of graphene and fabrication of energy storage devices.

**References.**

(1) Xu, S.; Zhang, Y.; Cho, J.; Lee, J.; Huang, X.; Jia, L.; Fan, J. A.; Su, Y.; Su, J.; Zhang, H.; et al. Stretchable Batteries with Self-Similar Serpentine Interconnects and Integrated Wireless Recharging Systems. *Nat. Commun.* **2013**, *4* (1), 1543. https://doi.org/10.1038/ncomms2553.
(2) Kim, Y. Y.; Yang, T.-Y.; Suhonen, R.; Välimäki, M.; Maaninen, T.; Kemppainen, A.; Jeon, N. J.; Seo, J. Photovoltaic Devices: Gravure-Printed Flexible Perovskite Solar Cells: Toward Roll-to-Roll Manufacturing (Adv. Sci. 7/2019). *Adv. Sci.* **2019**, *6* (7), 1970044. https://doi.org/10.1002/advs.201970044.
(3) Gates, B. D. Flexible Electronics. *Science (80-. ).* **2009**, *323* (5921), 1566 LP – 1567. https://doi.org/10.1126/science.1171230.
(4) Bauer, S. Sophisticated Skin. *Nat. Mater.* **2013**, *12* (10), 871–872. https://doi.org/10.1038/nmat3759.
(5) Wang, X.; Lu, X.; Liu, B.; Chen, D.; Tong, Y.; Shen, G. Flexible Energy-Storage Devices: Design Consideration and Recent Progress. *Adv. Mater.* **2014**, *26* (28), 4763–4782. https://doi.org/10.1002/adma.201400910.
(6) Stoller, M. D.; Park, S.; Zhu, Y.; An, J.; Ruoff, R. S. Graphene-Based Ultracapacitors. *Nano Lett.* **2008**, *8* (10), 3498–3502. https://doi.org/10.1021/nl802558y.
(7) Beidaghi, M.; Gogotsi, Y. Capacitive Energy Storage in Micro-Scale Devices: Recent Advances in Design and Fabrication of Micro-Supercapacitors. *Energy Environ. Sci.* **2014**, *7* (3), 867–884. https://doi.org/10.1039/C3EE43526A.
(8) Zhou, G.; Li, F.; Cheng, H.-M. Progress in Flexible Lithium Batteries and Future Prospects. *Energy Environ. Sci.* **2014**, *7* (4), 1307–1338. https://doi.org/10.1039/C3EE43182G.
(9) Gaikwad, A. M.; Whiting, G. L.; Steingart, D. A.; Arias, A. C. Highly Flexible, Printed Alkaline Batteries Based on Mesh-Embedded Electrodes. *Adv. Mater.* **2011**, *23* (29), 3251–3255. https://doi.org/10.1002/adma.201100894.
(10) El-Kady, M. F.; Kaner, R. B. Scalable Fabrication of High-Power Graphene Micro-Supercapacitors for Flexible and on-Chip Energy Storage. *Nat. Commun.* **2013**, *4* (1), 1475. https://doi.org/10.1038/ncomms2446.
(11) Beidaghi, M.; Wang, C. Micro-Supercapacitors Based on Interdigital Electrodes of Reduced Graphene Oxide and Carbon Nanotube Composites with Ultrahigh Power Handling Performance. *Adv. Funct. Mater.* **2012**, *22* (21), 4501–4510.

https://doi.org/10.1002/adfm.201201292.
(12) Liu, Y.; Zhang, B.; Xu, Q.; Hou, Y.; Seyedin, S.; Qin, S.; Wallace, G. G.; Beirne, S.; Razal, J. M.; Chen, J. Development of Graphene Oxide/Polyaniline Inks for High Performance Flexible Microsupercapacitors via Extrusion Printing. *Adv. Funct. Mater.* **2018**, *28* (21), 1706592. https://doi.org/10.1002/adfm.201706592.
(13) Wang, Y.; Shi, Z.; Huang, Y.; Ma, Y.; Wang, C.; Chen, M.; Chen, Y. Supercapacitor Devices Based on Graphene Materials. *J. Phys. Chem. C* **2009**, *113* (30), 13103–13107. https://doi.org/10.1021/jp902214f.
(14) Zeiger, M.; Jäckel, N.; Mochalin, V. N.; Presser, V. Review: Carbon Onions for Electrochemical Energy Storage. *J. Mater. Chem. A* **2016**, *4* (9), 3172–3196. https://doi.org/10.1039/C5TA08295A.
(15) Xiong, Z.; Liao, C.; Han, W.; Wang, X. Mechanically Tough Large-Area Hierarchical Porous Graphene Films for High-Performance Flexible Supercapacitor Applications. *Adv. Mater.* **2015**, *27* (30), 4469–4475. https://doi.org/10.1002/adma.201501983.
(16) Huang, P.; Lethien, C.; Pinaud, S.; Brousse, K.; Laloo, R.; Turq, V.; Respaud, M.; Demortière, A.; Daffos, B.; Taberna, P. L.; et al. On-Chip and Freestanding Elastic Carbon Films for Micro-Supercapacitors. *Science (80-. ).* **2016**, *351* (6274), 691 LP – 695. https://doi.org/10.1126/science.aad3345.
(17) Zhang, C. (John); McKeon, L.; Kremer, M. P.; Park, S.-H.; Ronan, O.; Seral-Ascaso, A.; Barwich, S.; Coileáin, C. Ó.; McEvoy, N.; Nerl, H. C.; et al. Additive-Free MXene Inks and Direct Printing of Micro-Supercapacitors. *Nat. Commun.* **2019**, *10* (1), 1795. https://doi.org/10.1038/s41467-019-09398-1.
(18) Acerce, M.; Voiry, D.; Chhowalla, M. Metallic 1T Phase MoS2 Nanosheets as Supercapacitor Electrode Materials. *Nat. Nanotechnol.* **2015**, *10* (4), 313–318. https://doi.org/10.1038/nnano.2015.40.
(19) Chen, H.; Müller, M. B.; Gilmore, K. J.; Wallace, G. G.; Li, D. Mechanically Strong, Electrically Conductive, and Biocompatible Graphene Paper. *Adv. Mater.* **2008**, *20* (18), 3557–3561. https://doi.org/10.1002/adma.200800757.
(20) Jung, S. M.; Mafra, D. L.; Lin, C.-T.; Jung, H. Y.; Kong, J. Controlled Porous Structures of Graphene Aerogels and Their Effect on Supercapacitor Performance. *Nanoscale* **2015**, *7* (10), 4386–4393. https://doi.org/10.1039/C4NR07564A.
(21) Yu, Z.; McInnis, M.; Calderon, J.; Seal, S.; Zhai, L.; Thomas, J. Functionalized Graphene Aerogel Composites for High-Performance Asymmetric Supercapacitors. *Nano Energy* **2015**, *11*, 611–620. https://doi.org/https://doi.org/10.1016/j.nanoen.2014.11.030.
(22) Li, L.; Secor, E. B.; Chen, K.-S.; Zhu, J.; Liu, X.; Gao, T. Z.; Seo, J.-W. T.; Zhao, Y.; Hersam, M. C. High-Performance Solid-State Supercapacitors and Microsupercapacitors Derived from Printable Graphene Inks. *Adv. Energy Mater.* **2016**, *6* (20), 1600909. https://doi.org/10.1002/aenm.201600909.
(23) Hernandez, Y.; Nicolosi, V.; Lotya, M.; Blighe, F. M.; Sun, Z.; De, S.; McGovern, I. T.; Holland, B.; Byrne, M.; Gun'Ko, Y. K.; et al. High-Yield Production of Graphene by Liquid-Phase Exfoliation of Graphite. *Nat. Nanotechnol.* **2008**, *3* (9), 563–568. https://doi.org/10.1038/nnano.2008.215.
(24) Coleman, J. N. Liquid Exfoliation of Defect-Free Graphene. *Acc. Chem. Res.* **2013**, *46* (1), 14–22. https://doi.org/10.1021/ar300009f.
(25) Paton, K. R.; Varrla, E.; Backes, C.; Smith, R. J.; Khan, U.; O'Neill, A.; Boland, C.; Lotya, M.; Istrate, O. M.; King, P.; et al. Scalable Production of Large Quantities of Defect-Free Few-Layer Graphene by Shear Exfoliation in Liquids. *Nat. Mater.* **2014**, *13* (6), 624–630. https://doi.org/10.1038/nmat3944.
(26) Nepal, A.; Singh, G. P.; Flanders, B. N.; Sorensen, C. M. One-Step Synthesis of Graphene via Catalyst-Free Gas-Phase Hydrocarbon Detonation. *Nanotechnology* **2013**, *24* (24), 245602. https://doi.org/10.1088/0957-4484/24/24/245602.
(27) Coleman, J. N. Liquid-Phase Exfoliation of Nanotubes and Graphene. *Adv. Funct. Mater.* **2009**, *19* (23), 3680–3695. https://doi.org/10.1002/adfm.200901640.
(28) Secor, E. B.; Ahn, B. Y.; Gao, T. Z.; Lewis, J. A.; Hersam, M. C. Rapid and Versatile Photonic Annealing of Graphene Inks for Flexible Printed Electronics. *Adv. Mater.* **2015**, *27* (42), 6683–6688. https://doi.org/10.1002/adma.201502866.
(29) Pastorova, I.; Botto, R. E.; Arisz, P. W.; Boon, J. J. Cellulose Char Structure: A Combined Analytical Py-GC-MS, FTIR, and NMR Study. *Carbohydr. Res.* **1994**, *262* (1), 27–47. https://doi.org/https://doi.org/10.1016/0008-6215(94)84003-2.
(30) Keiluweit, M.; Nico, P. S.; Johnson, M. G.; Kleber, M. Dynamic Molecular Structure of Plant Biomass-Derived Black Carbon (Biochar). *Environ. Sci. Technol.* **2010**, *44* (4), 1247–1253. https://doi.org/10.1021/es9031419.
(31) Khan, U.; O'Neill, A.; Lotya, M.; De, S.; Coleman, J. N. High-Concentration Solvent Exfoliation of Graphene. *Small* **2010**, *6* (7), 864–871. https://doi.org/10.1002/smll.200902066.
(32) Lotya, M.; Hernandez, Y.; King, P. J.; Smith, R. J.; Nicolosi, V.; Karlsson, L. S.; Blighe, F. M.; De, S.; Wang, Z.; McGovern, I. T.; et al. Liquid Phase Production of Graphene by Exfoliation of Graphite in Surfactant/Water Solutions. *J. Am. Chem. Soc.* **2009**, *131* (10), 3611–3620. https://doi.org/10.1021/ja807449u.
(33) Pech, D.; Brunet, M.; Durou, H.; Huang, P.; Mochalin, V.; Gogotsi, Y.; Taberna, P.-L.; Simon, P. Ultrahigh-Power Micrometre-Sized Supercapacitors Based on Onion-like Carbon. *Nat. Nanotechnol.* **2010**, *5* (9), 651–654. https://doi.org/10.1038/nnano.2010.162.
(34) Nemanich, R. J.; Solin, S. A. First- and Second-Order Raman Scattering from Finite-Size Crystals of Graphite. *Phys. Rev. B* **1979**, *20* (2), 392–401. https://doi.org/10.1103/PhysRevB.20.392.
(35) L.G., C.; Pimenta, M. A.; Neves, B. R. A.; Dantas, M. S. S.; Jorio, A. Influence of the Atomic Structure on the Raman Spectra of Graphite Edges. *Phys. Rev. Lett.* **2004**, *93* (24), 247401. https://doi.org/10.1103/PhysRevLett.93.247401.
(36) Ye, Y.-S.; Chen, Y.-N.; Wang, J.-S.; Rick, J.; Huang, Y.-J.; Chang, F.-C.; Hwang, B.-J. Versatile Grafting Approaches to Functionalizing Individually Dispersed Graphene Nanosheets Using RAFT Polymerization and Click Chemistry. *Chem. Mater.* **2012**, *24* (15), 2987–2997. https://doi.org/10.1021/cm301345r.
(37) Webb, M. J.; Palmgren, P.; Pal, P.; Karis, O.; Grennberg, H. A Simple Method to Produce Almost Perfect Graphene on Highly Oriented Pyrolytic Graphite. *Carbon N. Y.* **2011**, *49* (10), 3242–3249. https://doi.org/https://doi.org/10.1016/j.carbon.2011.03.050.